\documentclass[onecolumn, draft, 12pt]{IEEEtran}
\usepackage{amsmath,cases}
\usepackage{amsthm}
\usepackage{amssymb}
\usepackage{cite}
\usepackage{amsfonts}
\usepackage{enumerate}
\usepackage{url}
\usepackage[final]{graphicx}

\newcommand{\orderl}[1]{\leq_{\rm #1}}

\newcommand{\CDF}[2]{\ensuremath{F_{#1}\left(#2\right)}}
\newcommand{\pdf}[2]{\ensuremath{f_{#1}\left(#2 \right)}}
\newcommand{\vect}[1]{\ensuremath{\mathbf{#1}}}

\newcommand{\re}{\ensuremath{\mathbb{R}}}

\newcommand{\Pe}[2]{\ensuremath{P_{\rm e}^{\rm #1}\left(#2 \right)}} %Denotes P_e^(Desc)
\newcommand{\Peind}[3]{\ensuremath{P_{{\rm e}_{_{#2}}}^{#1}\left(#3 \right)}} % Denotes P_{e_{index}}^{var}
\newcommand{\AvgPe}[2]{\ensuremath{\overline{P}_{\rm {e}}^{\rm #1}\left(#2 \right)}}

\newcommand{\E}[2]{\ensuremath{\mathbb{E}_{#1}\left[ #2 \right]}}

\newcommand{\Q}[2]{\ensuremath{\mathcal{Q}^{#1}\left( #2 \right)}}
\newcommand{\vX}[1]{\vect{#1}}
\newcommand{\cx}{$X_{m}=x_{m}$ for $m = 1,\ldots,M$}

\newcommand{\wrt}{with respect to}

\newcommand{\F}{\mathfrak{G}}
\newcommand{\rv}{random variable}
\newcommand{\D}{\; d}

\newcommand{\g}{\ensuremath{g}}
\newcommand{\gfn}[2]{\ensuremath{g_{_{\rm #1}}\left( #2 \right)}}
\newcommand{\Sfunc}[1]{\ensuremath{S^{-}\left(#1 \right)}}
\newcommand{\LT}[2]{ \E{}{\exp(-#1 #2)}}
\newcommand{\gMod}[1]{\alpha_{{\rm #1}}}
\newcommand{\Xbold}[1]{\ensuremath{\mathbf{X}_{#1}}}
\newcommand{\Ybold}[1]{\ensuremath{\mathbf{Y}_{#1}}}
\newcommand{\MPSK}{$M$-PSK}
\newcommand{\MQAM}{$M$-QAM}
\newcommand{\cm}{c.m.}
\newcommand{\cmd}{c.m.d.}
\newcommand{\cerg}[1]{\E{#1}{C\left( \rho #1 \right)}}
\newcommand{\coa}[1]{\overline{C}^{#1}_{{\rm OA}}(\rho)}
\newcommand{\cci}[1]{\overline{C}^{#1}_{{\rm CI}}(\rho)}

\newcommand{\effch}{effective channel}
\newtheorem{theorem}{Theorem}

\begin{document}

\title{Applications of Stochastic Ordering to Wireless Communications}
\author{Cihan Tepedelenlio\u{g}lu, \emph{Member, IEEE}, Adithya Rajan, Yuan Zhang \thanks{The authors are with the School of Electrical, Computer, and Energy Engineering, Arizona
State University, Tempe, AZ 85287, USA. (Email:
\{cihan, arajan2, yzhang93\}@asu.edu).} } 
\maketitle
\begin{abstract}
Stochastic orders are binary relations defined on probability
distributions which capture intuitive notions like being larger or
being more variable. This paper introduces stochastic ordering of
instantaneous SNRs of fading channels as a tool to compare the
performance of communication systems over different channels.
Stochastic orders unify existing performance metrics such as ergodic
capacity, and metrics based on error rate functions for commonly
used modulation schemes through their relation with convex, and
completely monotonic (c.m.) functions. Toward this goal, performance
metrics such as instantaneous error rates of M-QAM and M-PSK
modulations are shown to be c.m. functions of the instantaneous SNR,
while metrics such as the instantaneous capacity are seen to have a
completely monotonic derivative (c.m.d.). It is shown that the
commonly used parametric fading distributions for modeling line of
sight (LoS), exhibit a monotonicity in the LoS parameter with
respect to the stochastic Laplace transform order. Using stochastic
orders, average performance of systems involving multiple random
variables are compared over different channels, even when closed
form expressions for such averages are not tractable. These include
diversity combining schemes, relay networks, and signal detection
over fading channels with non-Gaussian additive noise, which are
investigated herein. Simulations are also provided to corroborate
our results.

\end{abstract}

%Paper Body
\section{Introduction}
\label{sec:Intro}
Given the vast number of wireless systems with different purposes operating over fading channels, it is of interest to know how to decide whether one communication channel is superior to another. The performance of such systems are quantified by averaging a metric (e.g. bit or symbol error rates, or channel capacity) over the distribution of the random channel. Very often, when one channel is better than another in terms of a particular metric, it is also better with respect to another metric. However, this is not always true. Traditionally, answering this question has relied on single parameter comparisons between channels using characteristics such as diversity order, ``amount of fading'', Ricean factor and others \cite{simon_alouini05}. These are parametric approaches that quantify how much fading the channel exhibits, but do not provide a unified framework to compare channels across many different performance metrics. In this work, we propose to use stochastic orders to address this issue. 

The theory of stochastic orders (or dominance) provides a comprehensive framework to compare two random variables (RVs) or vectors \cite{shakedbook94}. The simplest and most widely used stochastic order compares the cumulative distribution functions (CDF) of two RVs, which defines a partial order between pairs of RVs. When the RVs represent instantaneous SNRs in a fading environment, this corresponds to comparing the outage probabilities in a wireless communication context. There are many other stochastic orders that capture comparisons of RVs in terms of size, and variability. Different than the related majorization theory \cite{marshallbook80,palomarbook07}, which defines a partial order on deterministic vectors, stochastic orders apply to random  variables. Stochastic ordering has become an indispensable tool in many increasingly diverse areas of probability and applied statistics over the past sixty years. Examples of such areas include reliability theory \cite{belzunce04}, actuarial sciences \cite{mullerbook_02}, risk analysis \cite{caballe07}, economics \cite{quirk62}, and stochastic processes \cite{ross_book96}. However, the applications of this set of tools in physical layer wireless communications are surprisingly very few, although it has found numerous applications in communication networks (please see \cite{ross_book96}, \cite[Ch. 13-14]{shakedbook94} and references therein).

We now review the limited literature on the applications of stochastic orders in physical layer communications. Bounds on the per cell sum rate under arbitrary fading in the high SNR regime have been obtained using the aforementioned outage-based ``usual stochastic order'' in \cite{levy09}. Stochastic ordering has also been applied to obtain bounds on the outage probability in Bluetooth piconets under
Ricean fading in \cite{karnik00}. In \cite{baitarokh08}, the usual stochastic order is used to bound monotone performance metrics in Ricean fading environments with beam selection. Reference \cite{tse09} shows that stochastic ordering of the SNR between the sender and any two receivers is sufficient for the existence of a degraded channel in a layered erasure broadcast channel modeled using the binary expansion model.

To the best of our knowledge, there is no systematic exploitation of the general stochastic ordering theory which can be used to provide a means for comparing wireless systems. All the above references use the usual stochastic order, which can be interpreted as a comparison of the outage probabilities, and do not exploit the full gamut of stochastic orders available \cite{shakedbook94}. In this paper, we give a wide range of examples illustrating how different stochastic orders are appropriate for comparing systems using different metrics with analytical properties such as monotonicity, convexity, and complete monotonicity, which shed light into the connections between performance metrics such as error rates and ergodic capacity. Additionally, we find the conditions for the preservation of inequalities satisfied by the averages of performance metrics of individual systems, when multiple such systems are combined. These may be combinations in parallel, in series, or otherwise, as may be seen in relay networks. Such a study permits the comparison of performance of systems, even in settings where closed-form expressions are not tractable.

\section{Stochastic Ordering Preliminaries}
\label{sec:Preliminaries}

The literature on stochastic ordering, primarily in reliability theory and statistics, delineates numerous stochastic orders, many of which fall under the subclass of {\it integral stochastic orders}.
We begin with a short description of the theory of integral stochastic orders,
which can be found in \cite{mullerbook_02,shakedbook94}. 

Let $\F$ denote a class of real valued functions $g : \re^{+}  \rightarrow \re$, and $X$ and $Y$ be RVs with CDFs $\CDF{X}{\cdot}$ and $\CDF{Y}{\cdot}$ respectively. We define the integral stochastic order with respect to $\F$ as \cite{mullerbook_02}:
\begin{align}
\label{eqn:integral_st_order_def}
X \leq_{\F} Y \Longleftrightarrow \E{}{g(X)} \leq \E{}{g(Y)} \;,\; \forall g \in \F.
\end{align}
In this case, $\F$ is known as a generator of the order $\orderl{\F}$. A stochastic order can have more than one generator. For a given stochastic order, it is of interest to identify ``large'' generators which are useful in identifying the equivalence of two orders. The largest generator set of functions for a stochastic order contains all other generators, and is termed the {\it maximal generator} \cite{mullerbook_02}. It is also of interest to find ``small'' generators which specify necessary conditions for the ordering of two RVs. We now give three examples of integral stochastic orders by specifying the corresponding generator set of functions $\F$.

\subsection{Usual Stochastic Order}
 The usual stochastic order compares the magnitude of two RVs. In this case a small generator $\F$ is the set of all non-decreasing \emph{indicator functions}: $\F=\{g(x): g(x)=I[x>\rho],\rho \in \re\}$. From \eqref{eqn:integral_st_order_def} it follows that this order is equivalent to comparing the CDFs of the RVs. Formally, we write 
\begin{equation}
\label{eqn:usual_stochastic_defn}
X \orderl{st} Y \Longleftrightarrow \CDF{X}{x} \geq \CDF{Y}{x} \; \forall x \;.
\end{equation}
To see the interpretation of this in the context of wireless channels, consider two channels to be compared, with {\effch}s $X:=|h^{X}|^2$ and $Y:=|h^{Y}|^2$, where $h^{X}$ and $h^{Y}$ correspond to the complex channel gains of two wireless systems. The usual stochastic ordering
of $X$ and $Y$ is equivalent to comparing their corresponding outage probabilities for all outage thresholds. 
The maximal generator for the usual stochastic order is the set of all increasing functions \cite{mullerbook_02}.
As a result, with the choice $g(x) = x$ in \eqref{eqn:integral_st_order_def}, we obtain $\E{}{X} \leq \E{}{Y}$ whenever $X \orderl{Lt} Y$, which agrees with the intuition that a larger {\rv} must have a larger mean value. 
\subsection{Convex Order}
\label{subsec:convex}
In this case $\F$ is the set of all convex functions, and
the order is denoted as $X \orderl{cx} Y$. Since $g(x)=x$ and $g(x)=-x$ are both convex, from (\ref{eqn:integral_st_order_def}), we have $\E{}{X} = \E{}{Y}$ whenever $X$ and $Y$ are convex ordered. Therefore, convex ordering establishes that the RVs have the same mean and $X$ is ``less variable'' than $Y$. Clearly, in the fading context, this can be used to identify channels with ``less fading''. Since $\orderl{cx}$ is a measure of variability, one would expect that a degenerate RV is less in the convex sense than any other RV with the same mean. Indeed, this is the case: If $F_X(x)=I[x \geq \mu]$, then  $X \orderl{cx} Y$ for all RVs $Y$ with $\E{}{Y}=\mu$. So the degenerate RV has an absolute minimum dispersion, as measured by the convex order, which is a consequence of Jensen's inequality.
%, and also the reason why for convex (concave) metrics such as the error rate (capacity) are minimized (maximized) in the absence of fading.

Many performance metrics, such as channel capacity, error rates for different modulations \cite{loyka10} and coding schemes in wireless systems are convex or concave functions of the instantaneous SNR. Therefore, establishing convex ordering of two RVs can help us qualitatively measure the relative average performance of the corresponding systems. Note that if instead of convex functions, the class $\F$ is chosen as the set of all concave functions, one would get the same order with a reversal in the inequality.

Verifying the usual stochastic ordering of two RVs is straightforward through \eqref{eqn:usual_stochastic_defn}. What follows are easily testable sufficient conditions for $X \orderl{cx} Y$. Let $\Sfunc{\g(x)}$ denote the number of sign changes of $\g(x)$ as $x$ increases over $[0,\infty)$, then $X \orderl{cx} Y$ if  $\E{}{X} = \E{}{Y}$ and any of the following hold \cite{shakedbook94}:
\begin{align}
   \label{eqn:cx_cond1}
   \Sfunc{\pdf{Y}{x} - \pdf{X}{x}} = 2 &\text{ and the sign sequence is }+, - , +.\\
   \label{eqn:cx_cond2}
   \Sfunc{\CDF{Y}{x} - \CDF{X}{x}} = 1 &\text{ and the sign sequence is }+, - \;,
\end{align}
where $\pdf{X}{\cdot}$ and $\pdf{Y}{\cdot}$ are the probability density functions (PDFs) of $X$ and $Y$ respectively.

Interestingly, to the best of our knowledge, although convex ordering of RVs is widely used in many other areas, it has never been used in physical-layer wireless communications.
\subsection{Laplace Transform Order}
\label{subsec:LTorder}
Similar to $\orderl{st}$ and $\orderl{cx}$, it is possible to order {\rv}s based on their Laplace transforms (LT). In this case, $\F = \lbrace \g(x) : \g(x) = - \exp\left( - \rho x \right) , \; \rho \geq 0 \rbrace$, so that
\begin{align}
\label{eqn:LT_expectations}
X \orderl{Lt} Y \Longleftrightarrow \E{}{\exp(-Y \rho)} \leq \E{}{\exp(-X \rho)} , \; \forall \; \rho > 0 \; .
\end{align}
Interpreting $\exp(-\rho x)$ as being proportional to the instantaneous error rate (as in the case for differential-PSK (DPSK) modulation and Chernoff bounds for other modulations), LT ordering of the instantaneous SNRs in \eqref{eqn:LT_expectations} can be interpreted as an inequality in the average error rates satisfied at all values of SNR $\rho$. One of the powerful consequences of LT ordering is that 
\begin{align}
\label{eqn:LT_order_relation}
X \orderl{Lt} Y \Longleftrightarrow \E{}{\g(X)} \geq \E{}{\g(Y)} \; ,
\end{align}
for all \emph{completely monotonic} ({\cm}) functions $g(\cdot)$ \cite[pp. 96]{shakedbook94}. A similar result with a reversal in the inequality states that 
\begin{align}
\label{eqn:LT_order_relation_1}
X \orderl{Lt} Y \Longleftrightarrow \E{}{\g(X)} \leq \E{}{\g(Y)} \;,
\end{align}
for all functions $\g(\cdot)$ that have a completely monotonic derivative (c.m.d) function. Recall that the derivatives of a {\cm} function alternate in sign. Also, they can be written as a positive mixture of decaying exponentials. More formally, a c.m. function by definition satisfies $(-1)^{n} \D^{n} \g(x)/dx^{n} \geq 0$, for $x >0$ and $n = 0,1,2,\ldots$, which by the celebrated \emph{Bernstein's theorem} is equivalent to the existence of a positive function $\mu(\cdot)$ such that $\g(x) = \int\limits_{0}^{\infty} \exp \left( -x\, u \right) \mu(u) \D u$ \cite[pp. 96]{shakedbook94}. It can be easily verified that {\cm} (c.m.d) functions are convex (concave) and decreasing (increasing). Further, if $g_{1}(x)$ is completely monotonic, and $g_{2}(x) \geq 0$ has a completely monotonic derivative, then the composition $g_{1}\left( g_{2}(x) \right)$ is completely monotonic. 

It is useful to mention that for any two RVs $X$ and $Y$, $X \orderl{cx} Y \Rightarrow Y \orderl{Lt} X$, which follows from the fact that $-\exp(-\rho x)$ is concave in $x$ for any $\rho >0$. Hence, convex ordering provides a method to generate or verify LT ordering between two RVs. Indeed either of the conditions \eqref{eqn:cx_cond1} or \eqref{eqn:cx_cond2} together with equal mean values for $X$ and $Y$ imply that $X \orderl{Lt} Y$. Further, observe that $X \orderl{st} Y \Rightarrow X \orderl{Lt} Y$, which follows since $-\exp(-\rho x)$ is increasing in $x$ for $\rho >0$.

In the rest of the paper, we illustrate the power of the stochastic ordering framework in comparing wireless channels and systems. We will investigate the convexity and complete monotonicity properties related to the  error rate and capacity expressions in Section \ref{sec:Metric_ordering}, which will facilitate comparing the average performances of systems by using \eqref{eqn:LT_order_relation} and \eqref{eqn:LT_order_relation_1}. In Section \ref{sec:order_distributions} we identify commonly used channel distributions which are LT or convex ordered. Section \ref{subsec:Multi_RV} investigates the conditions under which these stochastic orders are preserved in complex systems where the performance of their constituent parts satisfy an order. Finally, relevant simulations to supplement the theoretical results are provided in Section \ref{sec:simulations}.

\section{Ordering of Average Error Rate and Ergodic Capacity Metrics}
\label{sec:Metric_ordering}
\subsection{Symbol Error Rate}
\label{sec:SEP}
 It has been established in \cite{azizoglu02} that the error rate of binary signaling in the presence of noise with a uni-modal and differentiable PDF is a convex function of the SNR when maximum likelihood decoding is performed. Also, it is known that the instantaneous error probabilities of all one-dimensional and two-dimensional constellations with ML decoding in the presence additive white Gaussian noise (AWGN) is a convex function of the SNR \cite{loyka10}. In this section, we go one step further and establish the complete monotonicity of some two-dimensional modulation schemes, which will be useful in establishing inequalities between averaged performance metrics. It is well known that $\Q{}{\sqrt{x}}$ is {\cm} \cite{nesenbergs67}, from which the complete monotonicity of the instantaneous error rate of the form $\Pe{}{\rho x} = a \Q{}{\sqrt{b \rho x}}$ easily follows, for $a,b >0$. Here $a$ and $b$ are modulation dependent constants which can be chosen to get exact performance in some cases (e.g. $a=1$, $b=2$ for BPSK), or approximations in others ($a= 3/4, b= 4/5$ for $16$-QAM). For the exact case, it follows from \eqref{eqn:LT_order_relation} that $\E{}{\Pe{}{\rho Y}} \leq \E{}{\Pe{}{\rho X}}$, for $\rho > 0$ whenever $X \orderl{Lt} Y$.
 
We now establish, for the first time, the complete monotonicity of exact symbol error rates of square
{\MQAM} and {\MPSK} modulations which are not in the form $\Pe{}{\rho x}= a \Q{}{\sqrt{b \rho x}}$, and
offer sharper results than those mentioned above, since they do not rely on approximations.

The {\MPSK} symbol error rate is given by the following \cite[pp.195]{simon_alouini05}:
\begin{equation}
\label{eqn:MPSK_SEP}
\Pe{PSK}{\rho x} = \frac{1}{\pi}\int\limits_{0}^{(M-1)\pi/M}\exp\left( -\rho x \frac{\gMod{PSK}}{\sin^{2}\theta}\right) \D \theta \; ,
\end{equation}
where $\gMod{PSK} := \sin^{2}(\pi/M)$. After a change of variables, \eqref{eqn:MPSK_SEP} can be expressed as the Laplace transform of a positive function:
\begin{equation}
\label{eqn:MPSK_SEP_alt}
\Pe{PSK}{\rho x} = \frac{\sqrt{\gMod{PSK}}}{2 \pi}\int\limits_{0}^{\infty} e^{ - \rho u x} \frac{I \left[ u \geq \frac{\gMod{PSK}}{\sin^{2}\left((M-1)\frac{\pi}{M} \right)} \right]}{u\sqrt{u-\gMod{PSK}}} \D u \; ,
\end{equation}
which together with Bernstein's Theorem suggests that $\Pe{PSK}{\rho x}$ is {\cm}. 

Consider now the square {\MQAM} error rate function \cite[pp.195]{simon_alouini05}: 
\begin{equation}
\label{eqn:MQAM_SEP}
\Pe{QAM}{\rho x} = a \Q{}{\sqrt{\gMod{QAM}\rho x}} - b  \Q{2}{\sqrt{\gMod{QAM}\rho x}} \; ,
\end{equation}
where $\gMod{QAM} := 3/(M-1)$, $a := 4(\sqrt{M}-1)/\sqrt{M}$ and $b := a^{2}/4$. Note that $0 \leq b \leq a$. We claim that \eqref{eqn:MQAM_SEP} is \cm$\;$for any $a$, $b$ such that $b \leq a$. To see this, recall 
\begin{equation}
\label{eqn:Qfunc_generic}
\Q{k}{\sqrt{x}} = \frac{1}{\pi}\int\limits_{0}^{\pi/2k}\exp \left( -\frac{x}{2 \sin^{2} \theta} \right) \D \theta \; ,
\end{equation}
for $k =1,2$ \cite{simon_alouini05}. After a change of variables similar to \eqref{eqn:MPSK_SEP}, we obtain
\begin{align}
\label{eqn:MQAM_SEP_alt}
\Pe{QAM}{\rho x} = \frac{\sqrt{\gMod{QAM}}}{2 \pi}\int\limits_{0}^{\infty} e^{-u \rho x } \left[ \frac{a I\left[ 0.5 \leq u \leq 1 \right]}{u \sqrt{2u -1}} +  \frac{(a-b)I \left[ u \geq 1\right]}{u \sqrt{2u -1}} \right] \D u \;,
\end{align}
which is also {\cm} by Bernstein's theorem, since $b \leq a$. In conclusion, whenever $X \orderl{Lt} Y$, $\E{}{\Pe{}{\rho Y}} \leq \E{}{\Pe{}{\rho X}}$ for all average SNR $\rho$, where $\Pe{}{\cdot}$ could be given by either \eqref{eqn:MPSK_SEP} or \eqref{eqn:MQAM_SEP}. This follows from the definition of the LT order and the {\cm} properties of instantaneous error rates of {\MQAM} or {\MPSK} modulations. 

\subsection{Ergodic Capacity}
\label{subsec:ergodic_cap}
We now show that stochastic ordering of instantaneous {\effch} distributions implies that their ergodic channel capacities satisfy a corresponding inequality at all average SNRs. We begin with the case where only the receiver has channel status information (CSI). 

\subsubsection{Ergodic Capacity with Receive CSI only}
 The instantaneous capacity, conditioned on the {\effch} $X=x$ when only the receiver has CSI is given by $C(\rho x) = \log(1+\rho x)$, where $\rho$ is the average SNR. Since $d C(\rho x)/dx = \rho / (1+\rho x)$ is  {\cm} in $x$, from \eqref{eqn:LT_order_relation_1} $X \orderl{Lt} Y$ implies that the ergodic capacities satisfy $\cerg{X} \leq \cerg{Y}$ for $\rho \geq 0$. Recall that by \eqref{eqn:LT_expectations}, LT ordering of the channels $X$ and $Y$ can be interpreted as a comparison of the average error rates, when the instantaneous error rate is a decaying exponential. As a result, one can loosely say that {\it if the average error rates of two channels $X$ and $Y$ satisfy the inequality \eqref{eqn:LT_expectations} at all SNRs, then so do the ergodic capacities}. Surprisingly, however, the converse is not true, as we now illustrate. Consider a Pareto-type distribution, which is appropriate for modeling the instantaneous SINR in the presence of interference \cite{pun07} :
\begin{align}
\label{eqn:pareto_poor}
\CDF{X}{z} = \frac{z^{\beta}}{(1+z^{\beta})} \, , z > 0, \;  \beta >0 \;.
\end{align}
Using integration by parts and simplifying, we obtain 
\begin{align}
\label{eqn:pareto_cap}
\cerg{X} = \int\limits_{0}^{\infty} \frac{\rho}{(1+\rho z)(1+ z^{\beta})} \D z \;.
\end{align}
Taking the derivative with respect to $\beta$, it is seen that $\cerg{X}$ is a decreasing function of $\beta$, for $\rho > 0$. This shows that for $\beta^{X} \leq \beta^{Y}$, $\cerg{X} \geq \cerg{Y}$ for $\rho > 0$. On the other hand, since $\CDF{X}{z} = z^{\beta} + o(z^{\beta})$ near $z = 0$, the average symbol error rate for an exponential instantaneous error rate function satisfies $\E{}{\exp(- \rho X)} = \left( G_{c}\; \rho \right)^{-\beta} + o\left(\rho^{-\beta} \right)$, where $G_{c}$ is the array gain and $\beta$ is the diversity order \cite{wangGiannakis03}. Hence, as $\beta$ increases, the high-SNR average error rate decreases, while the capacity also decreases at all SNR $\rho$! Interpreting the ergodic capacity as what is achievable by coding over an i.i.d. time-extension of the channel, we reach the conclusion that even though $Y$ offers more diversity than $X$ for an uncoded system, the i.i.d. extension of $X$ lends itself to more diversity than that of $Y$. To put it more simply, at high SNR, it is possible for one channel to be superior to another in terms of error rates in the absence of coding, while being inferior when the capacity achieving code is employed over both channels.

\subsubsection{Channel Inversion and Delay-Limited Capacity}
When CSI is available at the transmitter, it can be used for power adaptation. A simple, suboptimal approach is to ``invert'' the channel at the transmitter, so that effectively the receiver sees a non-fading AWGN channel. Such an approach is viable only when $\E{}{X^{-1}} < \infty$, leading to a finite average transmit power. This is the case whenever the channel offers a diversity order strictly greater than one. Channel inversion has the advantage that a channel code designed for the AWGN channel can be used effectively, and that the code length need not depend on the channel coherence time to average out the fading. This ``delay-limited'' approach \cite{goldsmith02} gives rise to an achievable rate given by
\begin{align}
\label{eqn:C_CI}
\cci{X} = \log \left( 1+ \frac{\rho}{\E{}{X^{-1}}} \right) \;.
\end{align}
Clearly, since $g(x) = x^{-1}$ is a {\cm} function of $x$, $\E{}{X^{-1}} \geq \E{}{Y^{-1}}$, whenever $X \orderl{Lt} Y$. This implies $\cci{X} \leq \cci{Y}$ for all $\rho$, since $\cci{X}$ in \eqref{eqn:C_CI} is a decreasing function of  $\E{}{X^{-1}}$.

\subsubsection{Optimal Power and Rate Adaptation (OA)}
 It is well known that CI is not optimal, since when the channel gain becomes arbitrarily small, the transmitter uses extremely high power. To overcome this limitation, the \emph{optimal power and rate adaptation} scheme is proposed in \cite{goldsmith02}, where water-filling across time is performed subject to an average transmit power constraint. The capacity so obtained over a channel with instantaneous SNR $X$ is given by \cite{goldsmith02} :
\begin{align}
\label{eqn:cap_oa}
\coa{X} = \int\limits_{z_{t}(\rho)}^{\infty} \log \left( \frac{z_{t}(\rho)}{z}\right) \D \left[1-\CDF{X}{z} \right] \; ,
\end{align}
where $z_{t}(\rho)$ is the signaling threshold, which is implicitly governed by the power constraint as follows:
\begin{align}
\label{eqn:coa_ip_constr}
\int\limits_{z_{t}(\rho)}^{\infty} \left(\frac{1}{z_{t}(\rho)}-\frac{1}{z}\right)\D \CDF{X}{z} = \rho \; .
\end{align} 
It can be shown that $X \orderl{Lt} Y$ does not guarantee $\coa{X} \leq \coa{Y}$ for all $\rho$. However, in what follows, we will show that under the stronger assumption $X \orderl{st} Y$, $\coa{X} \leq \coa{Y}$ for all $\rho$.

Using integration by parts on \eqref{eqn:coa_ip_constr}, it is observed that for $X \orderl{st} Y$, we have $z_{t}^{X}(\rho) \geq z_{t}^{Y} (\rho)$. Now, integrating \eqref{eqn:cap_oa} by parts, under the assumptions that $\lim\limits_{z \to \infty}(1- \CDF{X}{z})\log(z/z_{t}^{X}(\rho)) = 0$ and $\lim\limits_{z \to \infty}(1- \CDF{Y}{z})\log(z/z_{t}^{Y}(\rho)) = 0$, it is seen that $\coa{X} \leq \coa{Y}$ for $\rho \geq 0$, since $z_{t}^{X}(\rho) \geq z_{t}^{Y}(\rho)$. Therefore, $X \orderl{st} Y \; \Rightarrow \coa{X} \leq \coa{Y}$, for $\rho >0$. 

\section{Ordering of Parametric Fading Distributions}
\label{sec:order_distributions}
We now proceed to show that commonly used parametric fading distributions are completely monotonic in the line of sight parameter with respect to LT and convex orders. 
\subsection{Nakagami Fading}
\label{subsubsec:Nakagami} 
 Consider Nakagami fading model, where the envelope $\sqrt{X}$ is Nakagami and the effective channel $X$ is Gamma distributed \cite{tega04}, with PDF given by 
\begin{align}
\label{eqn:gamma_PDF} 
\pdf{X}{x} = \frac{m^{m}}{\Gamma(m)}x^{m-1}\exp(-mx) \; , x \geq 0 \;.
\end{align}
Since $\E{}{\exp(- \rho X)} = \left( 1 + \rho / m \right)^{-m}$ is a decreasing function of $m$ for each $\rho$, it follows that if the $m$ parameters of two channel distributions satisfy
$m^{X} \leq m^{Y}$, then $X \orderl{Lt} Y$, where $X$ and $Y$ have normalized Gamma distributions with parameters $m^{X}$ and $m^{Y}$ respectively. This shows that for example, all the performance metrics in \eqref{eqn:MPSK_SEP} or \eqref{eqn:MQAM_SEP} that are {\cm} have averages over fading distributions that satisfy the inequality $\E{Y}{\Pe{}{\rho Y}} \leq \E{X}{\Pe{}{\rho X}}$ for all values of average SNR $\rho$. A similar claim with a reversal in the inequality can be made for the ergodic capacity metric. Note that the PDFs of $X$ and $Y$ in \eqref{eqn:gamma_PDF} are defined to satisfy $\E{}{X} =\E{}{Y}= 1$, independent of the fading parameter $m$. Hence, the improvements in error rate or ergodic capacity at all values of $\rho$ with increased $m$ is \emph{not} due to an improvement in average SNR. A stronger convex ordering result can also be established. Since $\E{}{X} = \E{}{Y}$, $m^{X} \leq m^{Y} \Rightarrow Y \orderl{cx} X$ can be shown by using \eqref{eqn:cx_cond1}. We can summarize the results herein by using the terminology that the normalized Gamma distribution is monotonically increasing in $m$ with respect to the orders $\orderl{Lt}$ and $\orderl{cx}$.
\subsection{Ricean Fading}
\label{subsubsec:Ricean_discussion}
 As in the Nakagami case, the Rice distribution will also be shown to be monotonic in the LoS parameter $K$ with respect to the orders $\orderl{Lt}$ and $\orderl{cx}$. The instantaneous SNR distribution is given by  
\begin{equation}
\label{eqn:Ricean_PDF} 
 \pdf{X}{x} =  (1+K)\exp(-K)\exp\left[-(K+1)x\right]I_{0}\left(2 \sqrt{K(K+1)x} \right) \; ,
\end{equation}
where $I_{0}(\cdot)$ is the modified Bessel function of the first kind of order zero. Clearly, $\E{}{X} = 1$ is independent of $K$. Taking the Laplace transform of \eqref{eqn:Ricean_PDF}, we have
 $\LT{X}{\rho} = (1+K)/(1+K+\rho)\exp \left[K \rho/(1 + K + \rho)\right]$, which decreases with $K$ for each $\rho$. This implies that, similar to the Nakagami case, increasing $K$ without increasing the average SNR improves the average symbol error rate, ergodic capacity, or any average metric obtained from a {\cm} or {\cmd} function. Thus, if $K^{X} \leq K^{Y}$ are the Ricean parameters of two channels, then $X \orderl{Lt} Y$. Similar to the Nakagami case, equation \eqref{eqn:cx_cond1} can be used to establish a stronger claim that $Y \orderl{cx} X$.

In this specific Ricean context, similar results for the ergodic capacity are found in \cite{lapidoth05_1} and the references therein, in a more general MIMO setting. However, in these results, either the channel power increases with an increase in the LoS component, or only an asymptotically large number of antennas is considered. 

\section{Communication Systems Involving Multiple RVs}
\label{subsec:Multi_RV}
In the following discussion, we will consider systems involving multiple independent random channel coefficients and compare their performance in two different sets of channels, where the {\effch}s associated with the first system are denoted by $\vect{X} :=[X_{1},\ldots,X_{M}]$ while those of the second channel by $\vect{Y} :=[Y_{1},\ldots,Y_{M}]$. Toward this goal, we use the following result \cite[pp. 97]{shakedbook94}, which shows that LT ordering is preserved by multivariate functions that are c.m.d.:
\begin{theorem}
\label{thm:closure}
Let $X_{1},\ldots , X_{M}$ be independent and $Y_{1},\ldots, Y_{M}$ also be independent. If $X_{m} \orderl{Lt} Y_{m} \; \text{ for}\; m = 1,\ldots,M$, then $g\left(X_{1},\ldots,X_{M}\right) \orderl{Lt} g\left(Y_{1},\ldots,Y_{M}\right)$ for all functions $g : \re^{m} \rightarrow \re^{+}$ such that for $m = 1,\ldots,M$, $\left(\partial/\partial x_{m}\right) g\left( x_{1},\ldots,x_{M}\right)$ is {\cm} in $x_{m}$, when all other variables are fixed.
\end{theorem}
We now investigate the systems for which the combined instantaneous SNR is given by a function $g(\vect{x}) := g(x_{1},\ldots,x_{M})$, which satisfies the conditions of Theorem \ref{thm:closure}. Unless otherwise mentioned, we will assume throughout that $X_{m} \orderl{Lt} Y_{m}$ for $m = 1,\ldots,M$.
 
\subsection{Maximum Ratio Combining}
\label{subsec:MRC}
Consider a SIMO diversity combining system with $M$ receive antennas which have complete CSI. If maximum ratio combining (MRC) is performed, conditioned on the {\effch}s {\cx}, the instantaneous SNR at the output of the combiner is proportional to  
\begin{equation}
\label{eqn:MRC_SNR}
g_{_{\rm MRC}}\left( \vX{x} \right) =  \sum\limits_{m=1}^{M} x_{m} \; ,
\end{equation}
which satisfies the conditions of Theorem \ref{thm:closure} as easily seen by taking derivatives. Thus, we infer that when MRC is performed, $g_{_{\rm MRC}}\left( \vect{X} \right) \orderl{Lt} g_{_{\rm MRC}}\left( \vect{Y} \right)$. Consequently, whenever  $\Pe{}{\rho x}$ is {\cm} and $C(\rho x)$ is {\cmd}, the average error rates satisfy $\E{\vect{Y}}{\Pe{}{\rho g_{_{\rm MRC}}\left( \vect{Y} \right) }} \leq  \E{\vect{X}}{\Pe{}{\rho g_{_{\rm MRC}}\left( \vect{X} \right) }}$ for all $\rho$, and the ergodic capacities satisfy $\E{\vect{Y}}{C(\rho g_{_{\rm MRC}}\left( \vect{Y} \right) )} \geq \E{\vect{X}}{C(\rho g_{_{\rm MRC}}\left( \vect{X} \right) )}$, for all $\rho$.

\subsection{Equal Gain Combining}
\label{subsec:EGC}
Next, assume that the SIMO diversity system adopts equal gain combining (EGC) at the receiver. In this case, conditioned on the instantaneous effective channels {\cx}, the instantaneous SNR at the combiner is proportional to 
\begin{equation}
\label{eqn:EGC_SNR}
g_{_{\rm EGC}} \left( \vX{x} \right) = \frac{1}{M}\left( \sum\limits_{m=1}^{M} \sqrt{ x_{m}} \right)^{2} \; .
\end{equation}
The derivative $\left(\partial/\partial x_{i}\right) g_{_{\rm EGC}} \left( x_{1},\ldots,x_{M}\right) = M^{-1}  \left(\sum\limits_{m=1}^{M} \sqrt{x_{m}} \right)/ \sqrt{x_{i}}$ is a {\cm} function of $x_{i}$, for $i = 1,\ldots,M$. Therefore, equal gain combining of a better set of branches results in a better system overall, as also expressed more rigorously after \eqref{eqn:MRC_SNR} in the MRC example.
\subsection{Selection Combining}
\label{subsec:SC}
In contrast to the previous two examples, this example shows that even though the individual branch instantaneous SNRs are LT ordered, the combined SNR at the output of the combiner need not be LT ordered.
For selection combining (SC), conditioning on the instantaneous {\effch}s {\cx}, we have 
\begin{equation}
\label{eqn:SC_SNR}
g_{_{\rm SC}}\left( \vX{x} \right) = \max_{m} x_{m} \; ,
\end{equation}
which is not differentiable, and hence is not {\cm}. In fact, $X_{m} \orderl{Lt} Y_{m}, \;  m = 1,\ldots,M$ does not imply $\max_{m} X_{m} \orderl{Lt} \max_{m} Y_{m}$. We provide a simple counterexample in Section \ref{sec:simulations}. This shows that even though channels $Y_{m}$ provide better average error rates at all $\rho$ than $X_{m}$, for $m = 1,\ldots,M$ for a SISO system, the composite SC channel does not.

\subsection{Multi-hop Amplify and Forward (AF)}
\label{subsec:MAF}
Consider a multi-hop system with $M$ links subject to AWGN, where $X_{m}$ is the {\effch} gain over the $m^{th}$ link. It is assumed that the $m^{th}$ node has channel information of the $(m-1)^{th}$ hop, for $m = 2,\ldots,M$, and the amplification factor for each node is the same. Conditioned on the instantaneous {\effch}s {\cx}, the SNR at the destination in this case is proportional to \cite{HasnaAlouini03}:
\begin{equation}
\label{eqn:AF_SNR}
g_{_{\rm MH-AF}}\left( \vX{x} \right) = \left[ \prod \limits_{m=1}^{M} \left( 1 + \frac{1}{x_{m}} \right)-1 \right]^{-1} \; .
\end{equation}
Taking the partial derivatives of $g_{_{\rm MH-AF}}\left( \vect{x} \right)$ with respect to each $x_{m}$ for $m = 1,\ldots,M$, it is seen that $g_{_{\rm MH-AF}}\left( \vect{x} \right)$ satisfies the conditions of Theorem \ref{thm:closure}. Thus, $g_{_{\rm MH-AF}}\left( \vect{X} \right) \orderl{Lt} g_{_{\rm MH-AF}}\left( \vect{Y} \right)$. As a result, the average error rates for the multi-hop AF system satisfy $\E{\vect{Y}}{\Pe{}{\rho \gfn{MH-AF}{\vect{Y}}}} \leq \E{\vect{X}}{\Pe{}{\rho \gfn{MH-AF}{\vect{X}}}}$, for $\rho >0$. Importantly, a closed-form expression for the average performance of this system is not tractable
for most practical channel distributions. Despite this, it is still possible to compare the error rates of two otherwise identical systems systems with two sets of LT ordered effective channels at all average SNRs.
\subsection{Multi-hop Channels with Decode and Forward}
\label{subsubsec:MDF}
Consider an $M$-hop channel, where each terminal decodes a received symbol into a binary alphabet and forwards it over to the next terminal. Let the instantaneous error rate over the $i^{th}$ link be given by $\Peind{}{i}{\rho x_{i}} , \; i = 1,\ldots,M$, where we assume $0 \leq \Peind{}{i}{x} \leq 1/2$ is {\cm}. For convenience, we define $\Xbold{1:m} := [X_{1}, \ldots , X_{m}]$ and let $\Peind{}{1:m}{\rho \Xbold{1:m}}$ be the combined instantaneous error rate of the first $1 \leq m \leq M$ hops. We have the following:
\begin{theorem}
\label{thm:DF_Pe_order}
Let $X_{1},\ldots,X_{M}$ be independent, and likewise for $Y_{1},\ldots,Y_{M}$. Suppose $X_{m} \orderl{Lt} Y_{m}$  for $m = 1,\ldots,M$. Then $\E{\Xbold{1:m}}{ \Peind{}{1:m}{\rho \Xbold{1:m}}} \geq \E{\Ybold{1:m}}{ \Peind{}{1:m}{\rho \Ybold{1:m}}} \; , \; m = 1,\ldots,M$. 
\end{theorem}
\begin{IEEEproof}
See Appendix.
\end{IEEEproof}

Note that Theorem \ref{thm:DF_Pe_order} and its proof carry over when each hop adopts $M$-ary modulation as well, provided that $\rho$ is large enough to ensure $0 \leq \E{\Xbold{1:m}}{ \Peind{}{1:m}{\rho \Xbold{1:m}}} \leq 1/2$. 
 
\subsection{Post Detection Combining}
\label{subsubsec:PDC}
Consider an $M$-antenna post-detection combining (PDC) scheme, where the instantaneous symbol error rate on the $m^{th}$ branch is $\Peind{}{m}{\rho x_{m}}$ and is {\cm} as in the previous example. The instantaneous probability of error of the PDC system is given by\footnote{We assume $M$ is odd. Extensions to even $M$ are straight-forward by adding a tie breaker term to \eqref{eqn:PDC_ber}.} \cite{adithya10} :
\begin{equation}
\label{eqn:PDC_ber}
\Peind{}{1:M}{\rho \Xbold{1:M}} = \sum\limits_{k = \frac{M+1}{2}}^{M}\sum\limits_{\mathcal{S}_{k}}\left( \prod\limits_{i \in \mathcal{S}_{k}} \Peind{}{i}{\rho x_{i}}\right)\left( \prod\limits_{j \in \mathcal{S}_{k}^{c}}\left( 1-\Peind{}{j}{\rho x_{j}} \right) \right) \; ,
\end{equation}
where $\mathcal{S}_{k}$ is a set running over all subsets of $\lbrace 1,\ldots,M \rbrace$ with $k$ elements. Taking expectation {\wrt} $\Xbold{1:M}$, which is assumed to have independent components, we have, 
\begin{equation}
\label{eqn:PDC_ber_1}
\E{\Xbold{1:m}}{ \Peind{}{1:m}{\rho \Xbold{1:m}}} = \sum\limits_{k = \frac{M+1}{2}}^{M}\sum\limits_{\mathcal{S}_{k}}\left( \prod\limits_{i \in \mathcal{S}_{k}}  \E{X_{i}}{\Peind{}{i}{\rho X_{i}}} \right)\left( \prod\limits_{j \in \mathcal{S}_{k}^{c}}\left( 1- \E{X_{j}}{\Peind{}{j}{\rho X_{j}}} \right) \right) \; .
\end{equation}
Clearly, the average error rate is an increasing function of any of the $\E{X_{m}}{\Peind{}{m}{\rho X_{m}}}$, since it is not possible to get improved performance by increasing the average error rate on any particular link. This shows that when $X_{m} \orderl{Lt} Y_{m}$, for $m = 1,\ldots,M$, and $\Peind{}{m}{\rho x}$ is {\cm}, so that the average error rates of PDC satisfy $ \E{X}{\Peind{}{m}{\rho X_{m}}} \geq  \E{Y}{\Peind{}{m}{\rho Y_{m}}}$ for $\rho > 0$, it follows that $ \E{\vect{X}}{\Peind{}{1:M}{\rho \Xbold{1:M}}} \geq \E{\vect{Y}}{\Peind{}{1:M}{\rho \Ybold{1:M}}}$ for $\rho >0$.
\subsection{Generalized Multi-branch Multi-hop AF Cooperative Relay Networks}
\label{sec:MB-MH_AF_CRN}
\begin{figure}[tb]
\begin{minipage}{1\textwidth}
\centering
\begin{center}
\includegraphics[scale=.5]{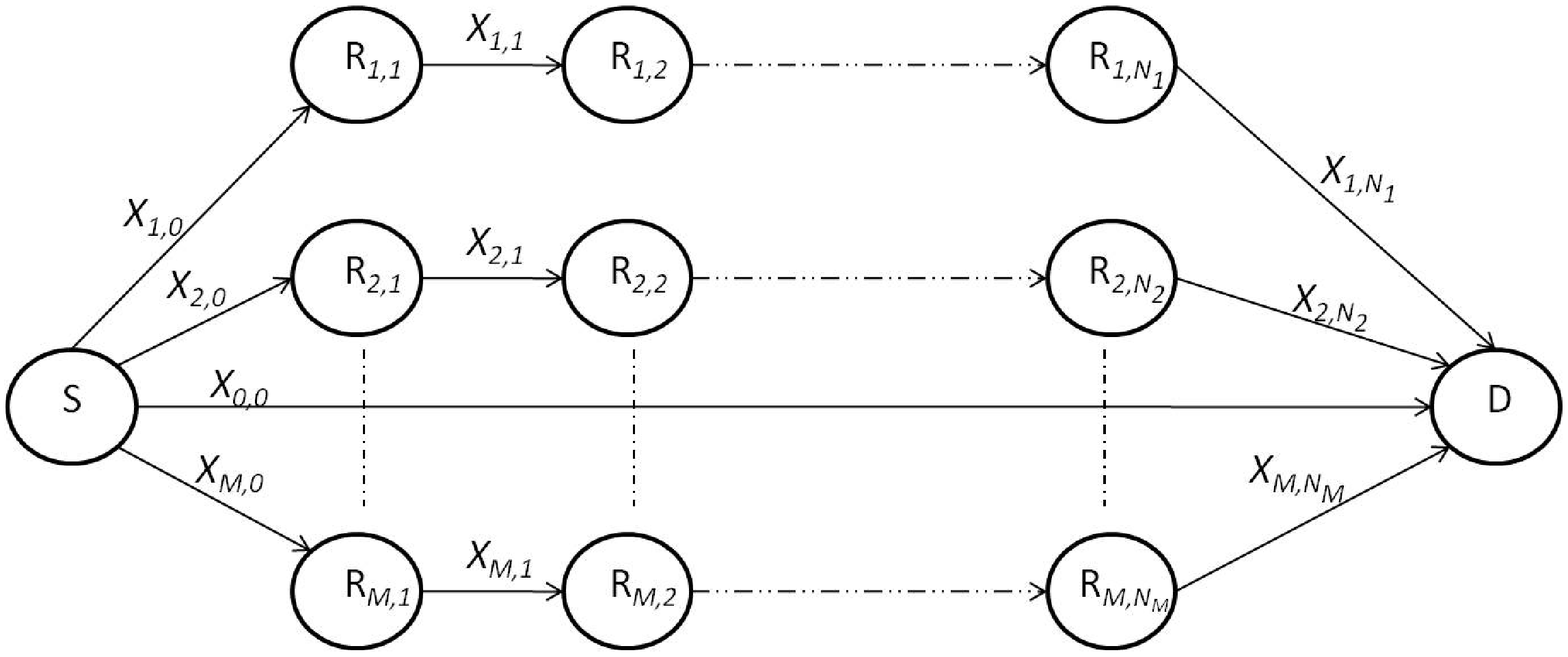}
 \caption{Multi-branch multi-hop cooperative relay network. ${\rm R}_{m,1} \ldots {\rm R}_{m,N_{_m}}$ represent the relays on the $m^{\rm th}$ link from the source ${\rm S}$ to the destination ${\rm D}$. The corresponding instantaneous {\effch} gains are denoted as $X_{m,0} \ldots X_{m,N_{_m}}$.}
\label{fig:MH-CD}
\end{center}
\end{minipage}
\end{figure}

We now consider the generalized relay structure illustrated Fig. \ref{fig:MH-CD}, which consists of $M$ independent branches, each involving $N_{m}$ relays, for $m = 1,\ldots, M$, which assist the direct link between the source ${\rm S}$ and the destination ${\rm D}$ by performing amplify and forward (AF). It is assumed that all the links are impaired by AWGN with fixed variance. This model requires the branches to communicate through mutually orthogonal channels, so that $M$ independent copies are available at the destination which performs MRC (using combining coefficients given in \cite{ribeiro05}). Although approximate expressions for the error rate have been obtained for the case of Ricean fading in \cite{ribeiro05}, closed-form expressions are intractable.  

\begin{figure}[tb]
\begin{minipage}{1\textwidth}
\centering
\begin{center}
\includegraphics[scale=.5]{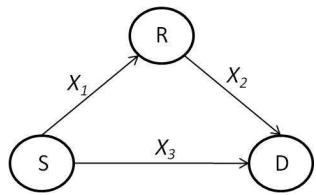}
 \caption{Two hop AF cooperative relay network.}
\label{fig:TwohopAF}
\end{center}
\end{minipage}
\end{figure}

Note that the two-hop fixed AF relay, which finds frequent application in cooperative diversity literature \cite{ribeiro05} and illustrated in Fig. \ref{fig:TwohopAF} is a special case of this general relay, with $M=1$ and $N_{m}=1$. Thus, the forthcoming results obtained for the general case apply for the two-hop relay as well.

We now show that the exact average symbol error rate can be compared over a number of fading distributions where the pairs of {\effch}s are LT ordered. To this end, we show that the output SNR of the MRC combiner at the destination satisfies the conditions of Theorem \ref{thm:closure}. Let $X_{0,0}$ denote the {\effch} on the direct link, and $X_{m,n}$ the {\effch} at the $n^{th}$ hop on link $m$. Since the destination performs MRC, the instantaneous output SNR is the sum of individual end-to-end branch SNRs, each of which are given by \eqref{eqn:AF_SNR}. Thus, conditioned on $X_{m,n}= x_{m,n}$ for $m=0,\ldots,M$ and $n = 0,\ldots,N_{m}$, and defining $\gfn{MB-MH-AF}{\vect{X}} := g_{_{\rm MB-MH-AF}}\left(x_{_{0,0}},x_{_{1,0}},\ldots , x_{_{1,N_{1}}},\ldots,x_{_{M,N_{M}}} \right) $, we have
\begin{align}
\label{eqn:MH_CD_linkSNR}
\gfn{MB-MH-AF}{\vect{X}} = \sum\limits_{m=1}^{M} \left[ \prod \limits_{n=1}^{N_{m}} \left( 1 + \frac{1}{x_{m,n}} \right)-1 \right]^{-1} +x_{0,0} \; .
\end{align}
As shown in the arguments following \eqref{eqn:AF_SNR}, the summand in the RHS of \eqref{eqn:MH_CD_linkSNR} has a {\cm} derivative in each variable. Combining this with Theorem \ref{thm:closure}, we have $\E{\vect{Y}}{\Pe{}{\rho \gfn{MB-MH-AF}{\vect{Y}}}} \leq \E{\vect{X}}{\Pe{}{\rho \gfn{MB-MH-AF}{\vect{X}}}}$ for $\rho >0$.

\subsection{Combined Multipath Fading and Shadowing}
\label{subsubsec:Multi_shadowing}
It is well known that the effect of shadow fading on the instantaneous SNR distribution can be modeled as a product of a shadowing {\rv} with a multipath fading {\rv} \cite{simon_alouini05}. Let $X_{1} \orderl{Lt} Y_{1}$ be the two multipath fading SNR distributions, and $X_{2} \orderl{Lt} Y_{2}$ be the two shadowing distributions. Then, from Theorem \ref{thm:closure}, it follows that the composite RV satisfies $X_{1}X_{2} \orderl{Lt} Y_{1}Y_{2}$, since $g(x_{1},x_{2}) = x_{1}x_{2}$ has a {\cm} derivative in each variable. We conclude that $\E{X_{1},X_{2}}{\Pe{}{\rho X_{1} X_{2}}} \geq \E{Y_{1},Y_{2}}{\Pe{}{\rho Y_{1} Y_{2}}} , \; \forall  \rho$, whenever $\Pe{}{\cdot}$ is {\cm}. Such conclusions can be drawn even in cases where the composite distribution of $X_{1}X_{2}$ or $Y_{1}Y_{2}$ cannot be written in closed-form.

\subsection{Systems with non-Gaussian Channel Noise}
\label{subsubsec:Non-Gauss}
In this discussion, we assume the following system model: 
\begin{equation}
\label{eqn:std_sys_model}
Z = \sqrt{\rho X}S + W \; ,
\end{equation}
where for simplicity, $S \in \lbrace -1,1 \rbrace$, $\rho X$ is the instantaneous SNR, $\rho$ the average SNR, and $W$ is non-Gaussian noise.
\subsubsection{Gaussian Mixture}
\label{subsubsec:GaussMix}
In this model, $W$ represents compound Gaussian noise (also called Gaussian mixture), which can be written as $W = \sqrt{A}G$, where $A$ is a positive valued RV, which represents the scale of $G$, and $G \sim \mathcal{N}(0,1)$. Such a formulation is possible for symmetric alpha-stable noise, Middleton class-A noise, as well as other compound Gaussian noise distributions. The error rate conditioned on the {\effch} $X=x$ is given by 
\begin{equation}
\label{eqn:Gaussian_mixture_BER}
\Pe{}{\rho x} = \E{A}{\Q{}{\sqrt{ \frac{2\rho x}{A}}}} \; ,
\end{equation}
which is a {\cm} function of $x$ as can be verified by differentiating inside the expectation with respect to $x$. Using \eqref{eqn:LT_order_relation}, this shows that when $X \orderl{Lt} Y$ then the average error rates satisfy $\E{X}{\Pe{}{\rho X}} \geq \E{Y}{\Pe{}{\rho Y}}$, even for mixed (compound) Gaussian noise.
Similar results can also be shown to hold for noise distributions such as the Laplace distribution which cannot be expressed as a compound Gaussian. 
\subsubsection{Bounded Noise}
\label{subsubsec:Bounded_noise}
Recall the system model from \eqref{eqn:std_sys_model}. If $|W| \leq C$ for some constant $C$, almost surely then $\CDF{W}{x} = 1$ for $x \geq C$ and $1 - \CDF{W}{\sqrt{2 x}}=0$ for $x^{2}/2 \geq C$. It is clear from Bernstein's theorem that a function, such as $1 - \CDF{W}{\sqrt{2 x}}$ with bounded support cannot be {\cm}. From this, we can conclude that if the noise is bounded, it is possible for two SNR distributions to be LT ordered, although $\E{Y}{\Pe{}{\rho Y}}$ need not be less than $\E{X}{\Pe{}{\rho X}}$ for all $\rho >0$. This negative result emphasizes the effect of the noise distribution in claims of ordering and concludes our discussion of systems with non-Gaussian noise.
\section{Simulations}
\label{sec:simulations}
We now corroborate our theoretical results using Monte-Carlo simulations. For ease of notation, we define $\AvgPe{X}{\rho} := \E{X}{\Pe{}{\rho X}}$ to denote the average error rates of SISO systems operated in the effective channel $X$. Also, we use $\AvgPe{\vect{X}}{\rho} := \E{\vect{X}}{\Pe{}{\rho \gfn{}{\vect{X}}}}$ to represent the average error rates of systems involving multiple effective channel coefficients.

\begin{figure}[tb]
\begin{minipage}{1\textwidth}
\centering
\begin{center}
\includegraphics[height=8.5cm,width=11.5cm]{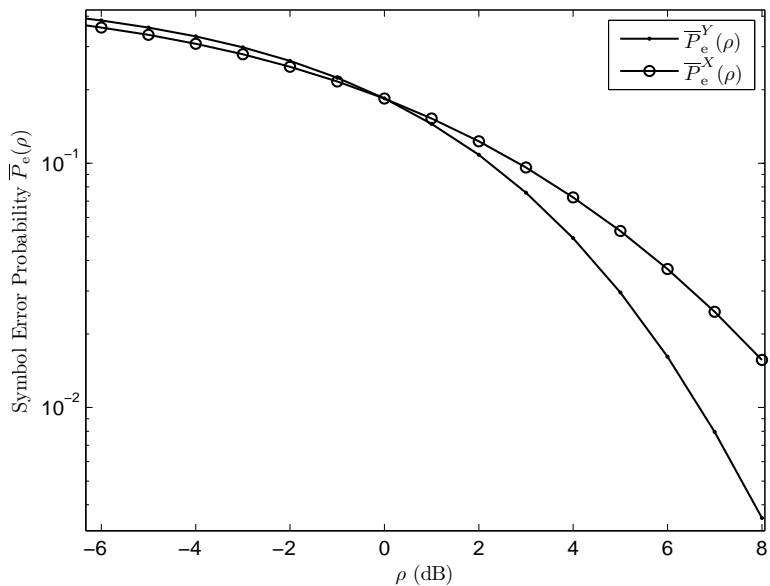}
\caption{Error probability comparison of DPSK modulation, under two different fading scenarios with LT ordered Pareto-type SNR distributions, using parameters $\beta^{X}=2$ and $\beta^{Y}=5$.}\label{fig:pareto_BER}
\end{center}
\end{minipage}
\end{figure}

\begin{figure}[tb]
\begin{minipage}{1\textwidth}
\centering
\begin{center}
\includegraphics[height=8.5cm,width=11.5cm]{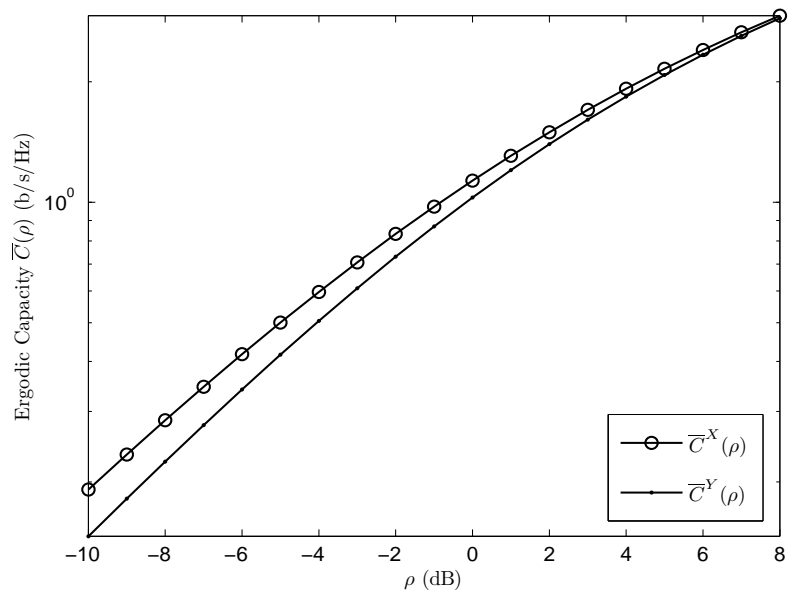}
\caption{Ergodic capacity comparison of two different fading scenarios with LT ordered Pareto-type SNR distributions, using parameters $\beta^{X}=2$ and $\beta^{Y}=5$. $\overline{C}^{X}(\rho)$ ($\overline{C}^{Y}(\rho)$) represents the ergodic capacity in the effective channel $X$ ($Y$).}\label{fig:pareto_cap}
\end{center}
\end{minipage}
\end{figure}

One of the central results of Section \ref{subsec:ergodic_cap} is that it is possible for one channel to be superior to another (in terms of error rates) at high SNR in the absence of coding, while being inferior when the capacity achieving code is used over both channels. This is illustrated in Fig. \ref{fig:pareto_BER}, which shows the comparative error rate performance of DPSK employed over an interference dominated fading channel with Pareto type distributed instantaneous SINR (having parameters  $\beta^{X}=2$ and $\beta^{Y}=5$. Clearly, since $\AvgPe{X}{\rho} < \AvgPe{Y}{\rho}$ for $\rho < -0.5$ dB and vice-versa for $\rho > -0.5$ dB, the system with effective channel $X$ is not better than that with effective channel $Y$ at every average SNR. On the other hand, Fig. \ref{fig:pareto_cap}, shows that the ergodic capacity of the system with instantaneous channel $X$ is consistently larger than that when operated in the channel $Y$ with parameter $\beta^{Y}=5$. 

\begin{figure}[tb]
\begin{minipage}{1\textwidth}
\centering
\begin{center}
\includegraphics[height=8.5cm,width=11.5cm]{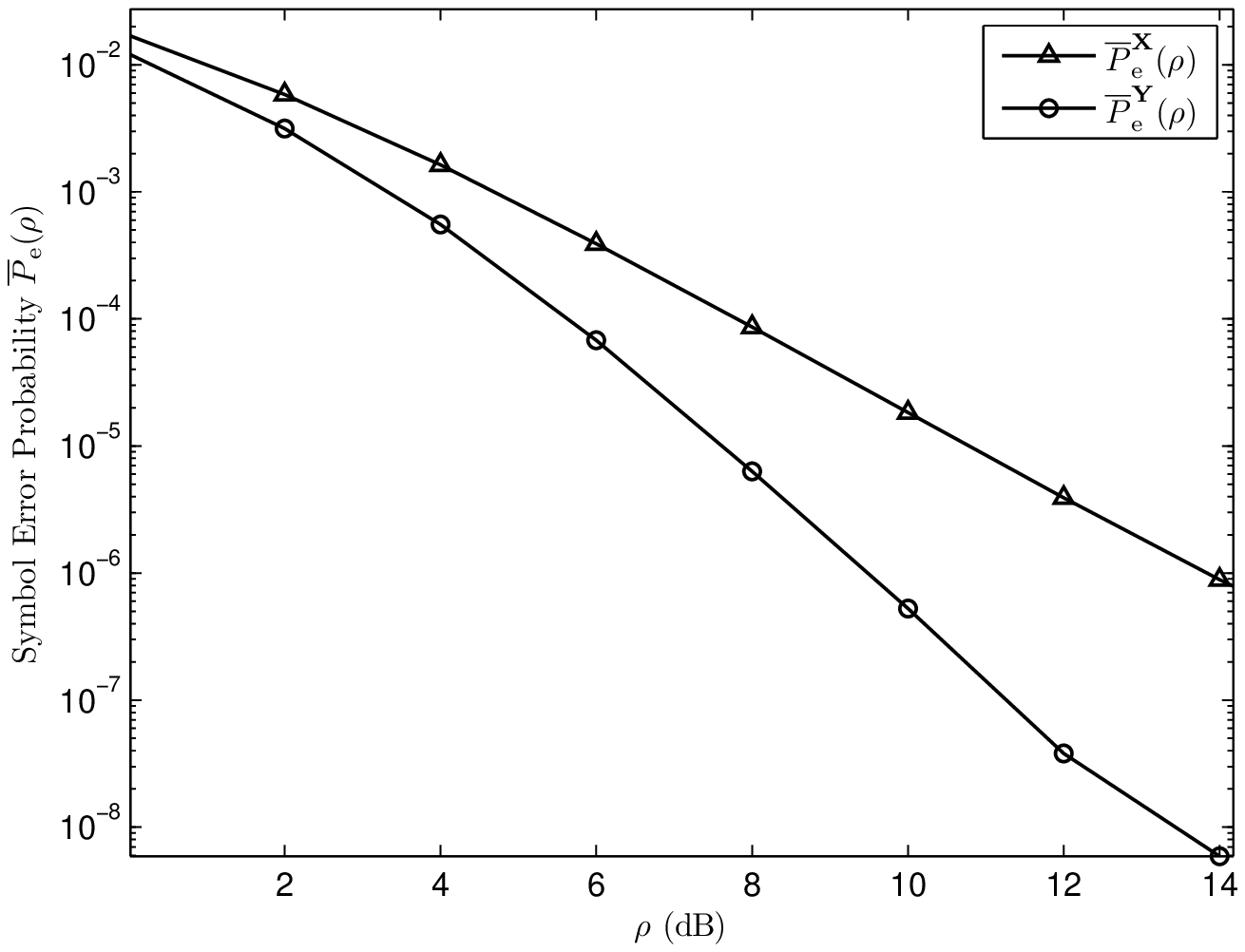}
\caption{Error rate comparison of maximum ratio combining using $L=3$ antennas with BPSK. $\AvgPe{\vect{X}}{\rho}$ corresponds to the average symbol error rate under Ricean fading with parameter $K^{X} = 2$ and $\AvgPe{\vect{Y}}{\rho}$ to the average symbol error rate under Ricean fading with parameter $K^{Y} = 5$.}\label{fig:Rice_BPSK_MRC_L3}
\end{center}
\end{minipage}
\end{figure}

\begin{figure}[tb]
\begin{minipage}{1\textwidth}
\centering
\begin{center}
\includegraphics[height=8.5cm,width=11.5cm]{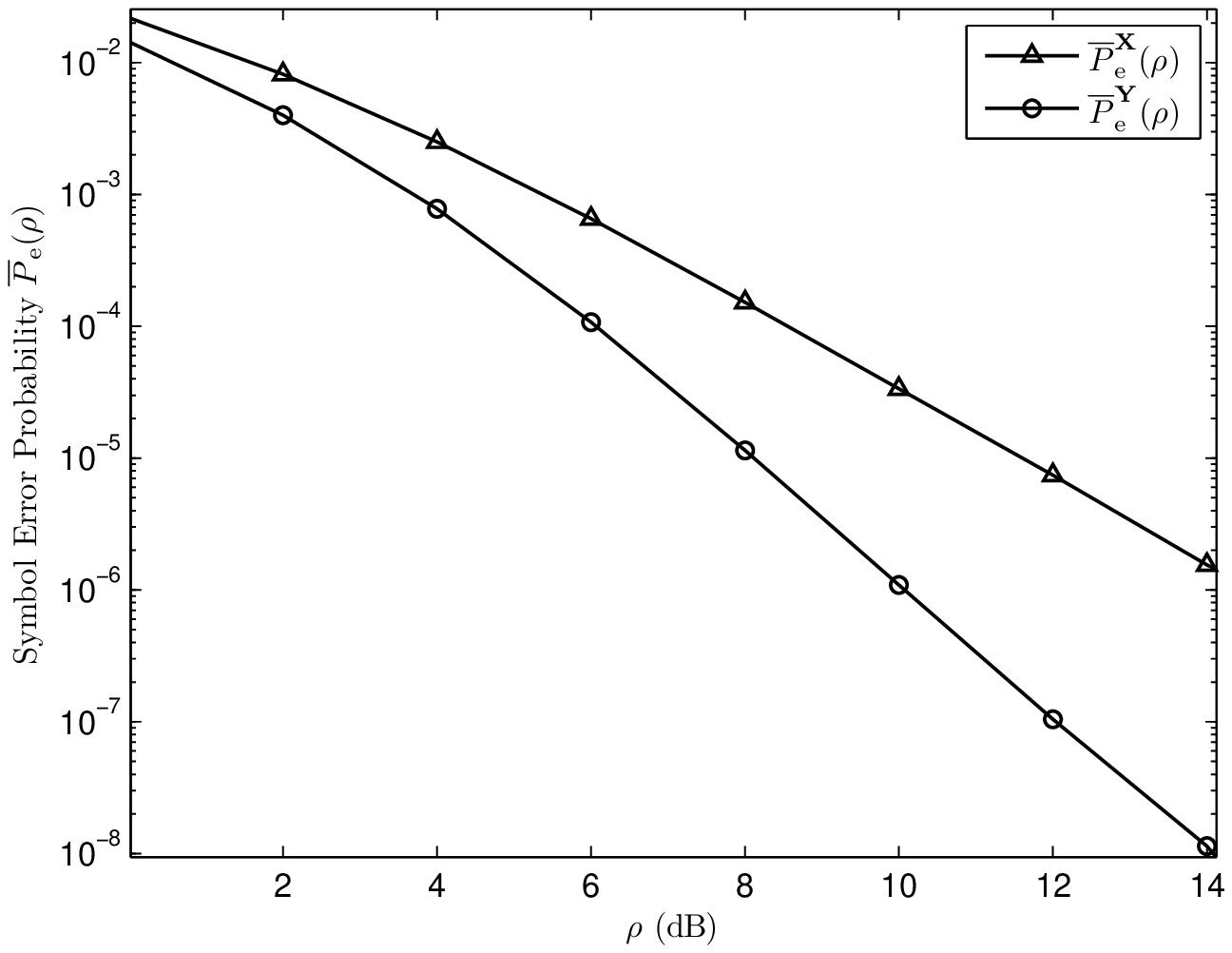}
\caption{Error rate comparison of equal gain combining using $L=3$ antennas with BPSK. $\AvgPe{\vect{X}}{\rho}$ corresponds to the average error rate under Ricean fading with parameter $K^{X} = 2$ and $\AvgPe{\vect{Y}}{\rho}$ to the average symbol error rate under Ricean fading with parameter $K^{Y} = 5$.}\label{fig:Rice_BPSK_EGC_L3}
\end{center}
\end{minipage}
\end{figure}

\begin{figure}[tb]
\begin{minipage}{1\textwidth}
\centering
\begin{center}
\includegraphics[height=8.5cm,width=11.5cm]{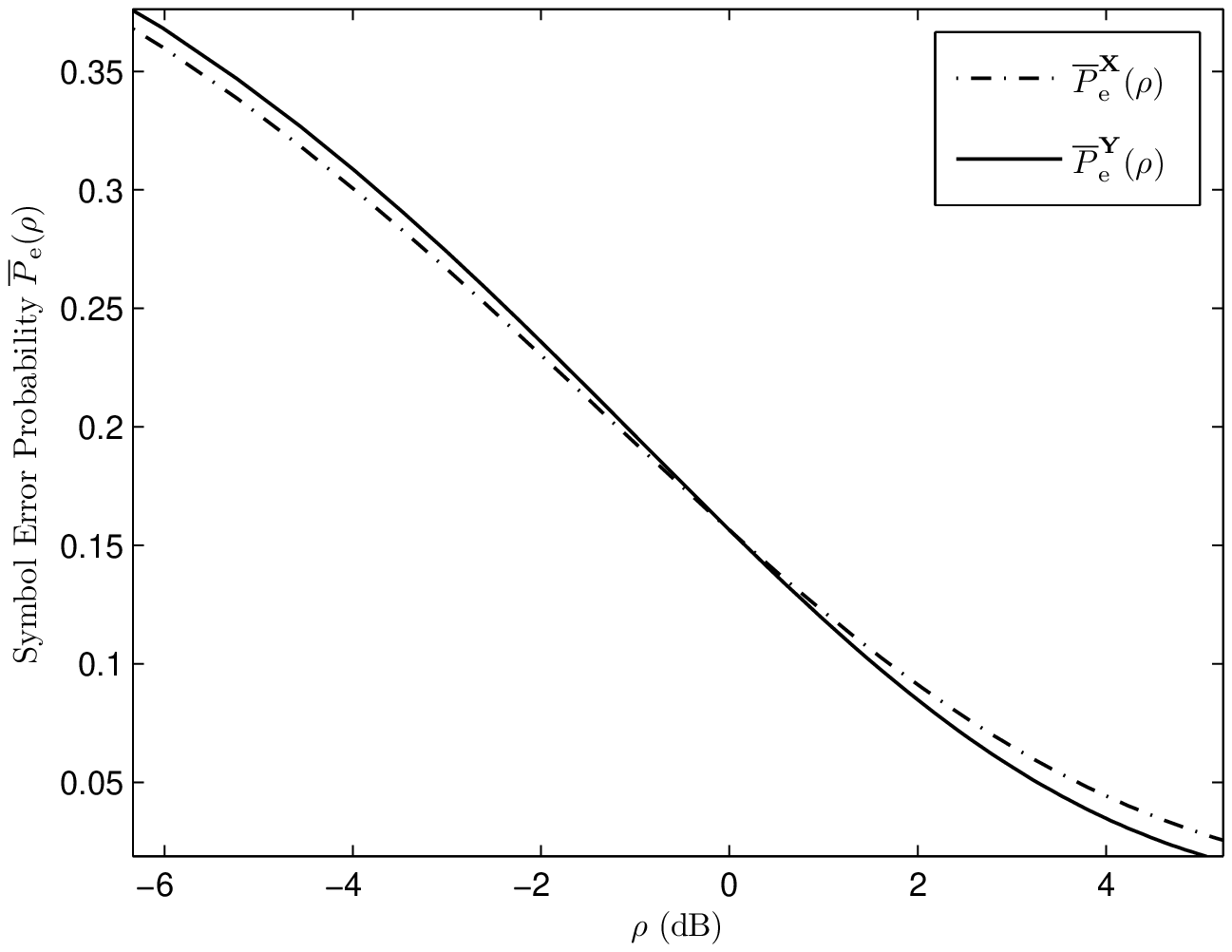}
\caption{Error rate comparison of selection combining using $L=3$ antennas with DPSK. $\AvgPe{\vect{X}}{\rho}$ corresponds to the average symbol error rate under Ricean fading with parameter $K^{X} = 2$ and $\AvgPe{\vect{Y}}{\rho}$ to the average symbol error rate under Ricean fading with parameter $K^{Y} = 5$.}\label{fig:SC_LT}
\end{center}
\end{minipage}
\end{figure}

Figures \ref{fig:Rice_BPSK_MRC_L3}, \ref{fig:Rice_BPSK_EGC_L3} and \ref{fig:SC_LT} show the performance of diversity combining schemes such as MRC, EGC and SC with $L=3$ branches over two sets of i.i.d Ricean fading channels with parameters $K^{X}=2$ and $K^{Y}=5$. Note that from Subsection \ref{subsubsec:Ricean_discussion}, $X_{m} \orderl{Lt} Y_{m}$ for $m=1,2,3$. The trend observed in the performance analysis curves obtained herein can be equivalently obtained for any other sets of LT ordered {\effch} random variables, using any modulation scheme whose error rate is a {\cm} function of the {\effch}.

In Fig. \ref{fig:Rice_BPSK_MRC_L3}, we demonstrate that LT ordering of the instantaneous SNR distributions for the individual branches can be used to compare average error rates when MRC is performed at the receiver. For $L=3$ receive diversity branches, it is observed that the error rate of BPSK in the channel with instantaneous SNR $\rho Y$ is consistently less than that in the channel with instantaneous SNR $\rho X$, which agrees with the fact that since the {\effch} for Ricean fading is {\cm} in $K$, $X_{m} \orderl{Lt} Y_{m}$, for $m = 1,2,3$, and hence $\AvgPe{\vect{Y}}{\rho} \leq \AvgPe{\vect{X}}{\rho}$ for $\rho > 0$.  

Figure \ref{fig:Rice_BPSK_EGC_L3} illustrates that when $X_{m} \orderl{Lt} Y_{m}$, for $m = 1,2,3$, we get $\AvgPe{\vect{Y}}{\rho} \leq \AvgPe{\vect{X}}{\rho}$ for $\rho > 0$ for the case of EGC employing BPSK. The error rate curves help demonstrate that fading channels with larger Ricean parameters offer smaller error rates than those with smaller Ricean parameters at all values of average SNR $\rho$ when EGC is used, as predicted in Subsection \ref{subsec:EGC}. Such a conclusion is not present in the literature due to the unavailability of a closed-form expression for the average error rate of coherent EGC in Ricean channels, which is applicable in all SNR regimes \cite{simon_alouini05}.

The comparative performance of SC using DPSK symbols is shown in Fig. \ref{fig:SC_LT}. It is evident that although the individual branch SNRs are LT ordered, $\AvgPe{\vect{Y}}{\rho} \geq \AvgPe{\vect{X}}{\rho}$, for $\rho < -0.4$ dB, while $\AvgPe{\vect{Y}}{\rho} \leq \AvgPe{\vect{X}}{\rho}$, for $ \rho \geq -0.4$ dB. This cross-over point is clearly depicted in Fig. \ref{fig:SC_LT} using a linear scale for the error rate axis, since it is more easily discernible compared to the conventional log scale. Hence, selection combining of a better set of channels (in terms of error rates) need not yield a better system overall, at low SNR.

\begin{figure}[tb]
\begin{minipage}{1\textwidth}
\centering
\begin{center}
\includegraphics[height=8.5cm,width=11.5cm]{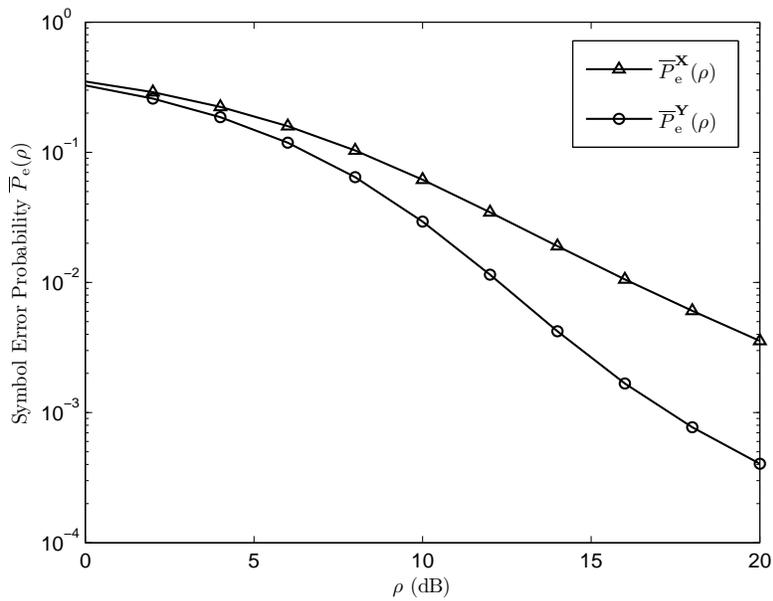}
\caption{Error rate comparison of $M=3$ hop amplify-forward relay with BPSK under Ricean fading. $\AvgPe{\vect{X}}{\rho}$ corresponds to the average symbol error rate under Ricean fading with parameter $K^{X} = 2$ and $\AvgPe{\vect{Y}}{\rho}$ to the average error rate under Ricean fading with parameter $K^{Y} = 5$.}\label{fig:MultihopAF_BER_BPSK}
\end{center}
\end{minipage}
\end{figure}

The performance of a multi-hop amplify and forward relay is studied in Fig. \ref{fig:MultihopAF_BER_BPSK}. We assume the model described in Section \ref{subsec:MAF} with $M=3$ relays under two different Ricean fading scenarios, one with parameter $K^{X}=2$ and the other with $K^{Y}=5$. It is observed that the average symbol error rate of $\vect{Y}$ is consistently less than that of $\vect{X}$ at all SNRs. This, due to the fact that $X_{m} \orderl{Lt} Y_{m} ,\; m = 1,2,3 \; \Rightarrow \AvgPe{\vect{Y}}{\rho} \leq \AvgPe{\vect{X}}{\rho}, \; \forall \rho$.

\begin{figure}[tb]
\begin{minipage}{1\textwidth}
\centering
\begin{center}
\includegraphics[height=8.5cm,width=11.5cm]{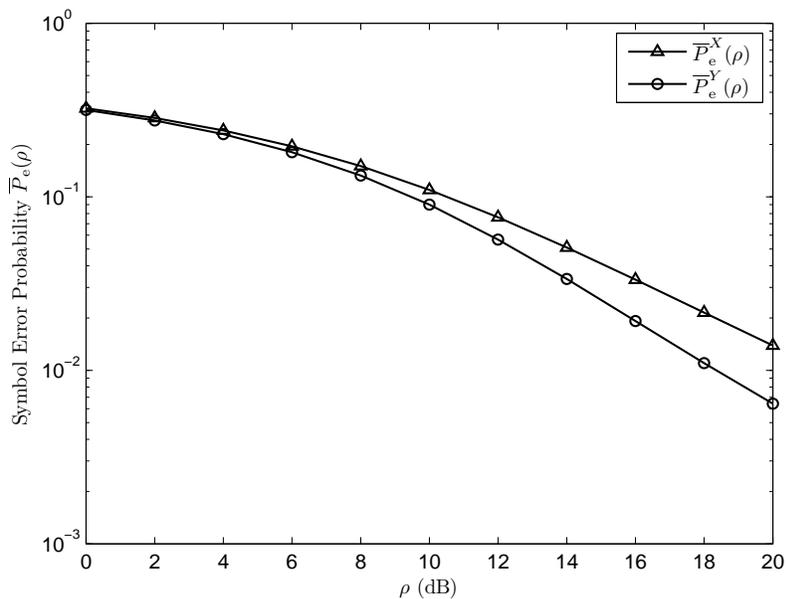}
\caption{Performance comparison of BPSK in compound Gaussian noise (normalized symmetric alpha-stable distribution with characteristic exponent $1.6$). $\AvgPe{X}{\rho}$ corresponds to the average symbol error rate under Ricean fading with parameter $K^{X} = 2$ and $\AvgPe{Y}{\rho}$ corresponds to the average symbol error rate under Ricean fading with parameter $K^{Y} = 5$.}
\label{fig:alphaStableBER}
\end{center}
\end{minipage}
\end{figure}

 Fig. \ref{fig:alphaStableBER} illustrates the comparative performance of an uncoded BPSK system over an additive compound Gaussian noise channel subject to two different Ricean fading effects modeled using parameters $K^{X}=2$ and $K^{Y}=5$. We show that $\AvgPe{Y}{\rho} \leq \AvgPe{X}{\rho}$ for all $\rho >0$, when the noise follows a symmetric alpha-stable distribution with a characteristic exponent of $1.6$. This shows that LT ordering results apply to systems with compound Gaussian noise, since an alpha-stable RV can be written as $\sqrt{A}G$, where $G \sim \mathcal{CN}(0,1)$ and $A$ is a positively skewed alpha-stable RV \cite{adithya10}. Such results are not found in literature, since a closed-form expression for the average error rate of BPSK under Ricean fading with symmetric alpha-stable noise is analytically intractable. In fact, even for the special case of $K=1$ i.e. Rayleigh fading, a closed-form expression valid in the asymptotic high SNR regime is known \cite{adithya10}.

\begin{figure}[tb]
\begin{minipage}{1\textwidth}
\centering
\begin{center}
\includegraphics[height=8.5cm,width=11.5cm]{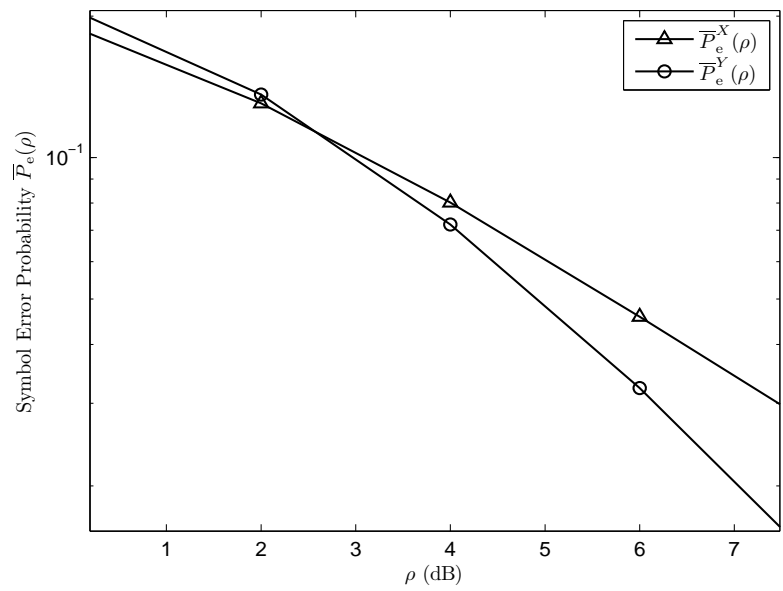}
\caption{Performance comparison of BPSK in noise with finite support (symmetric uniformly distributed noise with unit variance). $\AvgPe{X}{\rho}$ corresponds to the average symbol error rate under Ricean fading with parameter $K^{X} = 2$ and $\AvgPe{Y}{\rho}$ corresponds to the average error rate under Ricean fading with parameter $K^{Y} = 5$.}\label{fig:BoundedBER1}
\end{center}
\end{minipage}
\end{figure}

 In direct contrast to the results for the compound-Gaussian noise case, LT ordering of {\effch}s does not imply that the average error rate performance for noise with bounded support will satisfy the corresponding inequality at all SNR. In fact, as depicted in Fig. \ref{fig:BoundedBER1}, where the unit-variance noise is assumed to be uniformly distributed on $[-\sqrt{3},\sqrt{3}]$, it is observed that for $\rho < 2.6$ dB, $\AvgPe{X}{\rho} \leq \AvgPe{Y}{\rho}$, while the opposite holds for $\rho > 2.6$ dB. This corroborates the claim of Subsection \ref{subsubsec:Bounded_noise}, which states that LT ordering of {\effch}s does not imply that the average error rates satisfy $\AvgPe{Y}{\rho} \leq \AvgPe{X}{\rho}$ for all $\rho > 0$, under noise with finite support.

\section{Conclusions}
In this paper, we illustrate the power of stochastic orders such as the convex order and the LT order, which have never been used in physical layer communication/information theory, to relate and unify existing performance metrics such as ergodic capacity and error rate functions through their relationship with completely monotonic functions. We first identify that the instantaneous symbol error rate functions for various signaling constellations such as {\MPSK} and {\MQAM} are completely monotonic functions of the instantaneous SNR. Recognizing the importance of LT ordering of instantaneous SNR distributions, we identify parametric fading distributions such as Nakagami and Ricean, which are monotonic in the LoS parameters in the orders $\orderl{Lt}$ and $\orderl{cx}$. We also lay the groundwork to find the conditions for the preservation of inequalities satisfied by the averages of performance metrics of individual systems, when multiple such systems are combined, even when closed form expressions for such averages are not tractable. These include diversity combining schemes such as MRC, EGC and a variety of relay networks.

 In summary, this framework provides a novel approach to compare the performance of a vast range of systems on the basis of the analytical properties of the performance metric such as monotonicity, convexity, or complete monotonicity, even in settings where closed-form expressions are not tractable.

\appendix [Proof of Theorem \ref{thm:DF_Pe_order}]
\label{proof:DF_Pe_order}
For any $m$, viewing the $m$-hop channel as a series cascade of the first $m-1$ hops and the $m^{th}$ hop, we have the following:
\begin{equation}
\label{eqn:DF_proof_1}
\Peind{}{1:m}{\rho \Xbold{1:m}} = \Peind{}{1:m-1}{\rho \Xbold{1:m-1}}(1-\Peind{}{m}{\rho x_{m}}) + (1 - \Peind{}{1:m-1}{\rho \Xbold{1:m-1}})\Peind{}{m}{\rho x_{m}} \; ,
\end{equation}
for $m = 2,\ldots,M$. To prove the theorem, we will use induction. Clearly, Theorem \ref{thm:DF_Pe_order} holds for $m=1$. Taking expectation of both sides of \eqref{eqn:DF_proof_1}, we have
\begin{align}
\label{eqn:DF_proof_2}
\E{\Xbold{1:m}}{ \Peind{}{1:m}{\rho \Xbold{1:m}}} &= \E{\Xbold{1:m-1}}{ \Peind{}{1:m-1}{\rho \Xbold{1:m-1}}}\left( 1-\E{X_{m}}{ \Peind{}{m}{\rho X_{m}}} \right)\nonumber \\
&+ \left(1- \E{\Xbold{1:m-1}}{ \Peind{}{1:m-1}{\rho \Xbold{1:m-1}}} \right)\E{X_{m}}{ \Peind{}{m}{\rho X_{m}}} \; .
\end{align}
We have $\E{\Xbold{1:m-1}}{ \Peind{}{1:m-1}{\rho \Xbold{1:m-1}}} \geq \E{\Ybold{1:m-1}}{ \Peind{}{1:m-1}{\rho \Ybold{1:m-1}}}$ by the induction hypothesis, and $\E{\Xbold{1:m}}{ \Peind{}{1:m}{\rho \Xbold{1:m}}} \geq \E{\Ybold{1:m}}{ \Peind{}{1:m}{\rho \Ybold{1:m}}}$ follows because $\Peind{}{m}{\cdot}$ is {\cm} and $X_{m} \orderl{Lt} Y_{m}$. The theorem then follows because the RHS of \eqref{eqn:DF_proof_2} is of the form $P_{1}(1-P_{2}) + P_{2}(1-P_{1})$, which is an increasing function of both $P_{1}$ and $P_{2}$, since $0 \leq P_{1} \leq 1/2$, $0 \leq P_{2} \leq 1/2$.

\bibliographystyle{IEEEtran}
\nocite{*}
\bibliography{ST-Paper}
\newpage
%-----------------%

\end{document}